\documentclass[%
 reprint,
superscriptaddress,
 amsmath,amssymb,
 aps,
prb,citeautoscript
]{revtex4-2}

\usepackage{graphicx}
\usepackage{dcolumn}
\usepackage{bm}
\usepackage{lipsum}
\usepackage{siunitx}
\usepackage{placeins}
\usepackage{hyperref}

\usepackage{nameref}
\usepackage{amstext}
\usepackage{mathtools} 
\usepackage{amsmath,amssymb}    
\usepackage{booktabs}
\usepackage{array}
\usepackage{xcolor}
\usepackage{tabularx}
\usepackage{ulem}

\usepackage{booktabs}
\DeclareSIUnit{\rad}{rad}
\DeclareSIUnit{\deg}{deg}

\hypersetup{
        colorlinks   = true,
        citecolor    = blue,
        linkcolor    = blue}

\newcommand{\biscco}{$\text{Bi}_{2}\text{Sr}_{2}\text{CaCu}_{2}\text{O}_{8+d}$ }
\DeclareSIUnit\bar{bar}
\DeclareSIUnit\torr{Torr}

\makeatletter
\def\@email#1#2{%
 \endgroup
 \patchcmd{\titleblock@produce}
  {\frontmatter@RRAPformat}
  {\frontmatter@RRAPformat{\produce@RRAP{*#1\href{mailto:#2}{#2}}}\frontmatter@RRAPformat}
  {}{}
}%

\let\svthefootnote\thefootnote
\newcommand\freefootnote[1]{%
  \let\thefootnote\relax%
  \footnotetext{#1}%
  \let\thefootnote\svthefootnote%
}

\begin{document}

\preprint{APS/123-QED}

\title{%Controlling the Josephson coupling of twisted cuprate van der Waals heterostructures
Twisted cuprate van der Waals heterostructures with controlled Josephson coupling 
}

\author{Mickey Martini}
 \affiliation{Leibniz Institute for Solid State and Materials Science Dresden (IFW Dresden), 01069 Dresden, Germany}%Lines break automatically or can be forced with \\
 \affiliation{Institute of Applied Physics, Technische Universität Dresden, 01062 Dresden, Germany}

\author{Yejin Lee}
 \affiliation{Leibniz Institute for Solid State and Materials Science Dresden (IFW Dresden), 01069 Dresden, Germany}%Lines break automatically or can be forced with \\
 \affiliation{Institute of Applied Physics, Technische Universität Dresden, 01062 Dresden, Germany}
 
    \author{Tommaso Confalone}
 \affiliation{Leibniz Institute for Solid State and Materials Science Dresden (IFW Dresden), 01069 Dresden, Germany}%Lines break automatically or can be forced with \\
  \affiliation{Institute of Applied Physics, Technische Universität Dresden, 01062 Dresden, Germany}

 \author{Sanaz Shokri}
 \affiliation{Leibniz Institute for Solid State and Materials Science Dresden (IFW Dresden), 01069 Dresden, Germany}%Lines break automatically or can be forced with \\
 \affiliation{Institute of Applied Physics, Technische Universität Dresden, 01062 Dresden, Germany}
  
  \author{Christian~N.~Saggau}
 \affiliation{Leibniz Institute for Solid State and Materials Science Dresden (IFW Dresden), 01069 Dresden, Germany}%Lines break automatically or can be forced with \\

 \author{Daniel Wolf}
\affiliation{Leibniz Institute for Solid State and Materials Science Dresden (IFW Dresden), 01069 Dresden, Germany}%Lines break automatically or can be forced with \\
 
   \author{Genda Gu}
 \affiliation{Condensed Matter Physics and Materials Science Department, Brookhaven National Laboratory, Upton, NY 11973, USA}%Lines break automatically or can be forced with \\
 
   \author{Kenji Watanabe}
 \affiliation{Research Center for Functional Materials, National Institute for Materials Science, 1-1 Namiki, Tsukuba 305-0044, Japan}%Lines break automatically or can be forced with \\
 
    \author{Takashi Taniguchi}
\affiliation{International Center for Materials Nanoarchitectonics, National Institute for Materials Science,  1-1 Namiki, Tsukuba 305-0044, Japan
}%Lines break automatically or can be forced with \\

 \author{Domenico Montemurro}
 \affiliation{Department of Physics, University of Naples Federico II, 80125 Naples, Italy}%Lines break automatically or can be forced with \\    

  \author{Valerii\,M.\,Vinokur}
 \affiliation{Terra Quantum AG, Kornhausstrasse 25,
CH-9000 St.\,Gallen, Switzerland}
\affiliation{Physics Department, CUNY, City College of City University of New York, 160 Convent Ave, New York, NY 10031, USA}
  
 \author{Kornelius Nielsch}
 \affiliation{Leibniz Institute for Solid State and Materials Science Dresden (IFW Dresden), 01069 Dresden, Germany}%Lines break automatically or can be forced with \\
 \affiliation{Institute of Applied Physics, Technische Universität Dresden, 01062 Dresden, Germany}
 \affiliation{Institute of Materials Science, Technische Universität Dresden, 01062 Dresden, Germany}

 \author{Nicola Poccia}
% \author{Nicola Poccia${}^{*}$ }
 % \author{Nicola Poccia(\url{einstein@gmail.com})}
\altaffiliation{\url{n.poccia@ifw-dresden.de}}
 \affiliation{Leibniz Institute for Solid State and Materials Science Dresden (IFW Dresden), 01069 Dresden, Germany}%Lines break
% \email{n.poccia@ifw-dresden.de}

%\received{\today}
             
%\published{\today}

\begin{abstract}

Twisted van der Waals (vdW) heterostructures offer a unique platform for engineering the efficient Josephson coupling between cuprate thin crystals harboring the nodal superconducting order parameter. Preparing the vdW heterostructures-based Josephson junction comprising stacked cuprates requires maintaining an ordered interface with preserved surface superconductivity. Here, we report the preparation of the Josephson junction out of the stacked \biscco crystals using the cryogenic dry transfer technique and encapsulating the junction with an insulating layer, that protects the interface during the electrical contacts evaporation at the $\SI{1e-6}{\milli\bar}$ base pressure. We find that the Josephson critical current $I_{\text c}$ has a maximum at low twist angles, comparable to that of the bulk intrinsic Josephson junctions, and is reduced by two orders of magnitude at twist angles close to $\SI{45}{\degree}$. The reduction of $I_{\text c}$ occurs due to a mismatch between superconducting $d$-wave order parameters, which suppresses the direct Cooper pair tunneling.

\end{abstract}

\pacs{Valid PACS appear here}% PACS, the Physics and Astronomy
                             % Classification Scheme.

\maketitle

%The BSCCO thin crystals are an advantageous material for the technological Josephson junctions since even monolayers comprising a single BSCCO elementary cell remain superconducting\,\cite{yu2019high} holding the $d$-wave superconducting gap\,\cite{morikawa1999angle, basov2005electrodynamics}.

%BSCCO represents an ideal building block to reali

\section{Introduction}

Twisted van der Waals (vdW) heterostructures comprising thin cuprate superconductors have been proposed as an adaptable platform for Josephson junctions (JJs) with controllable properties\,\cite{kawabata2004macroscopic,yokoyama2007theory,volkov2021josephson, can2021probing, tummuru2022josephson, volkov2023magic}. These structures offer a unique opportunity for revealing an interplay between superconducting order parameter (SOP) symmetries in the superconducting crystals. The cuprate compound,\,\biscco (BSCCO) represents an ideal system for realizing the vdW heterostructures. Such a heterostructure forms a stack of the intrinsic JJs along the BSCCO crystallographic $c$-axis, comprising the superconducting CuO\textsubscript{2} bilayers sandwiched between the insulating [SrO-BiO] bilayers\,\cite{kleiner1992intrinsic}. Cleaving the BSCCO between the BiO planes forms atomically thin crystals with the superconducting transition temperature as high as that of bulk\,\cite{liao2018superconductor,yu2019high, zhao2019sign}, holding the $d$-wave superconducting gap\,\cite{morikawa1999angle, basov2005electrodynamics}. Because of the nodal gap in the $d$-wave SOP \cite{klemm1998angular, bille2001models}, the
Josephson tunneling across the twisted BSCCO junctions is expected to be suppressed at certain angles. Nevertheless, past experiments on the twisted BSCCO junctions did not detect any angular dependence of the Josephson critical current\,\cite{li1999bi, zhu2021presence, latyshev2004c}. This can be understood as a result of the high-temperature annealing in oxygen, which not only restores the surface superconductivity but also induces the undesired structural distortions at the junction. At the same time, the strong angular dependence of the critical current compatible with the $d$-wave pairing symmetry has been reported on the micron-thick whisker BSCCO junctions\,\cite{takano2002d}. The $d$-wave nature of the SOP has been also demonstrated in the grain boundary (GB)-based cuprate JJs and superconducting quantum interference devices fabricated using the bicrystal and tricrystal techniques\,\cite{mannhart2002experiments,hilgenkamp2002grain,lombardi2002intrinsic,tsuei2004robust}. The symmetry of the SOP was easy to observe in the GB junctions since the in-plane coherence length is much larger than the coherence length along the $c$-axis. Yet, the GBJJs have inevitable multifaceted interfaces in all three dimensions that create additional complexity in controlling Josephson coupling\,\cite{hilgenkamp1996implications}.

\begin{figure*}[t!]
    \center
    \includegraphics[width=1\textwidth]{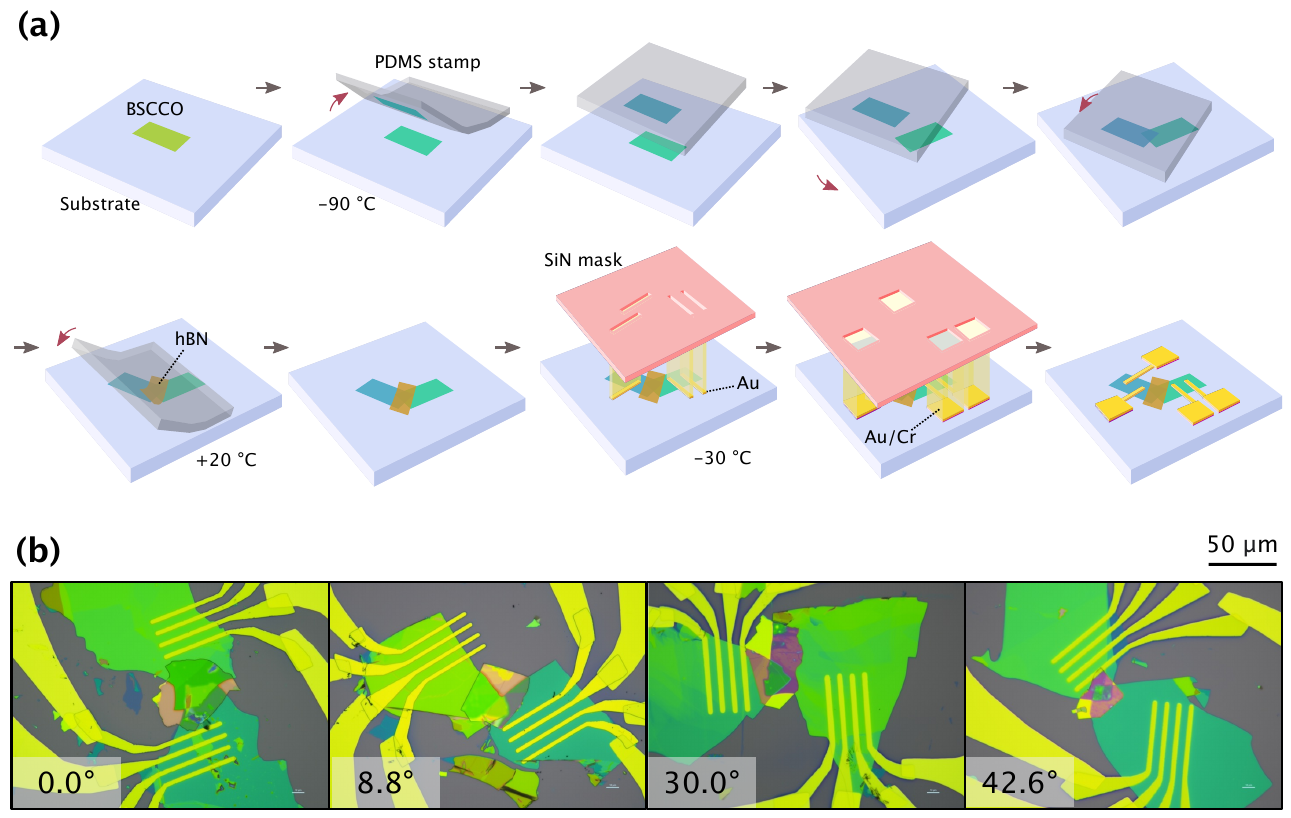}
    \caption{(a) Schematics of the twisted JJs fabrication process. (b) Optical micrographs of devices for JJs twisted at four different angles. The twist angle is specified on the bottom left of each subfigure.}
    \label{fig:stacking}
\end{figure*}

Furthermore, the small, $\le$0.1\,nm\,\cite{naughton1988orientational}, BSCCO $c$-axes coherence length, the high mobility oxygen dopants above $\SI{200}{\kelvin}$ \,\cite{poccia2011evolution, campi2015inhomogeneity}, and the chemical reactivity of the oxygen under an ambient atmosphere in the thin BSCCO crystals\,\cite{novoselov2005two, huang2015reliable} hinder the realization of vertical heterostructures with coherent interfaces free of detrimental disorder. An innovative cryogenic stacking technique in an inert atmosphere has recently enabled the atomically flat interface between twisted BSCCO crystals with preserved interfacial superconductivity\,\cite{zhao2021emergent, lee2021twisted}. In the resulting JJs, the critical current exhibited a strong angular dependence compatible with the $d$-wave SOP and an unusual non-monotonic temperature behavior due to the unconventional sign change of the nodal superconducting gap\,\cite{volkov2021josephson, tummuru2022josephson}. Around $\SI{45}{\degree}$ where the $d$-wave SOP of the two crystals mismatch maximally, the remaining superconducting coherence was ascribed to the presence of a finite second harmonics in the Josephson current-phase relation associated with co-tunneling of Cooper pairs across the junction\,\cite{volkov2021josephson, tummuru2022josephson}. The second harmonics was identified through the observation of fractional Shapiro steps and the analysis of the Fraunhofer patterns\,\cite{zhao2021emergent}. Time-Reversal Symmetry broken phases in twisted bilayers of cuprate high-temperature superconductors\,\cite{kuboki1996proximity, can2021probing,can2021high,tummuru2022twisted,liu2023making,volkov2023current} have been theoretically proposed, potentially leading to the chiral Majorana modes \,\cite{mercado2022high,margalit2022chiral}. 

Here, we build several JJs with different twist angles using the recently established dry and cryogenic stacking technique \cite{zhao2021emergent}. We encapsulate the junction region with an insulating crystal which enables us to prevent the detrimental effects of disorder during the device fabrication\,\cite{lee2022encapsulating}, especially during the evaporation of electrical contacts in high-vacuum. Since our chamber has a base pressure of $\SI{1e-6}{\milli\bar}$ and is not under the high-vacuum condition as in Ref.\,\cite{zhao2021emergent}, the reactivity of water molecules during the metal deposition is still too high. Therefore, an additional encapsulating layer is essential for preserving a pristine interface between the thin BSCCO crystals. By doing so, we also extend the lifetime of the device for several days.

\section{Fabrication of high-quality Josephson Junctions}

\begin{figure*}[t!]
    \center
    \includegraphics[width=.95\textwidth]{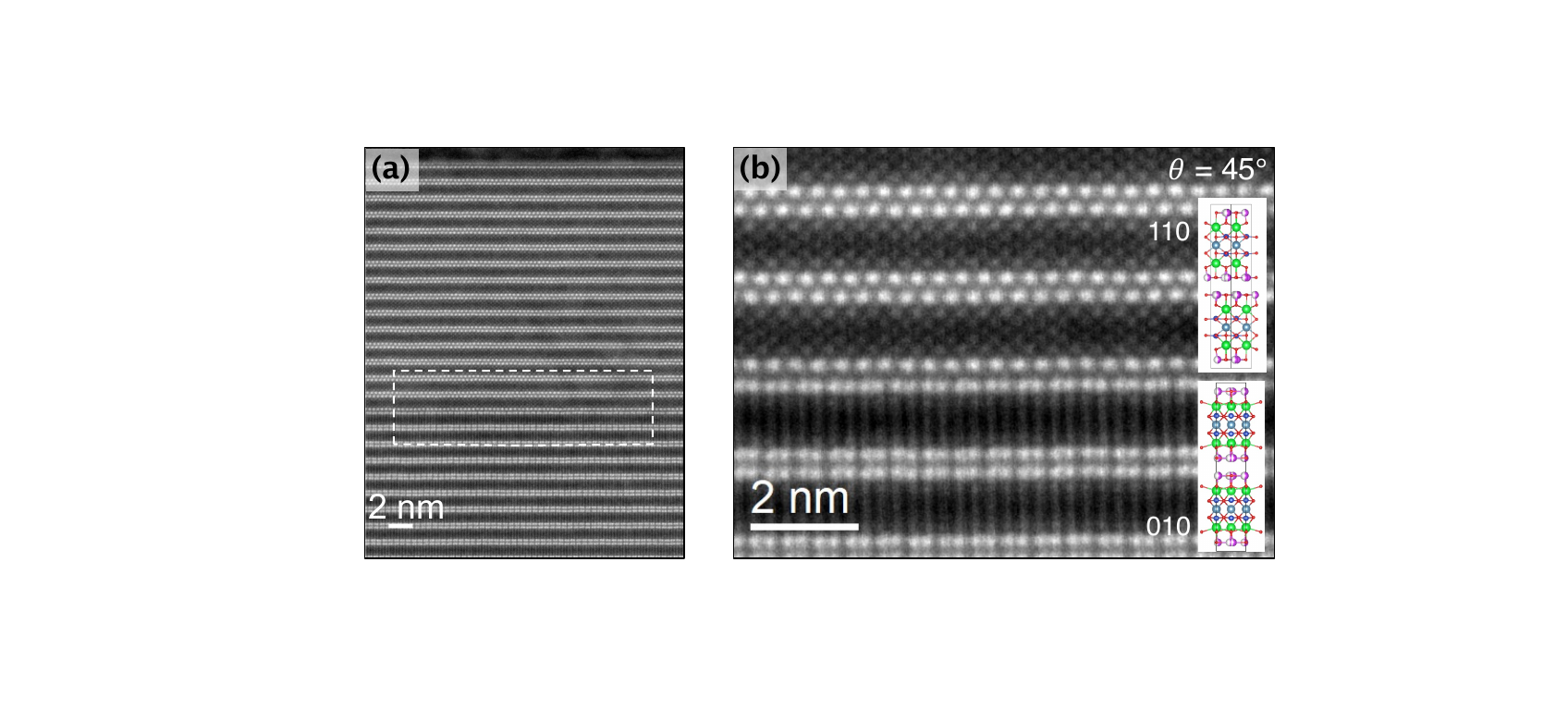}
    \caption{(a) High-angle annular dark-field scanning transmission electron microscopy (HAADF-STEM) image of a twist BSCCO junction at \SI{45}{\degree} displaying the junction's cross section between top and bottom crystals along $c$-axis. The brightest spots are identified with Bi atoms. (b) STEM image at higher magnification recorded across the interface within the dashed rectangle in (a). Inset: Illustration of crystalline order for both crystals in the heterostructure.}
    \label{fig:TEM}
\end{figure*}

We fabricate seven JJs devices consisting of two BSCCO crystals with a controlled twist angle using the cryogenic stacking technique in a glove box filled with argon. The twist angle $\theta$ ranges below $\SI{45}{\degree}$: two samples have $\theta = \SI{0}{\degree}$, two samples have twist angle close to $\SI{45}{\degree}$ ($\SI{42.6}{\degree}$ and $\SI{43.2}{\degree}$) and three samples are twisted at intermediate angles ($\theta = \SI{8.8}{\degree},\SI{21.5}{\degree}, \text{and}\,\, \SI{30.0}{\degree}$). The key steps of the JJ fabrication are sketched in Fig.\,\hyperref[fig:stacking]{1(a)}, and the details of the technique are presented in Ref.\,\cite{zhao2021emergent}. First, an optimally doped BSCCO flake is mechanically exfoliated using the scotch tape on SiO\textsubscript{2}/Si, previously treated with the oxygen plasma to enhance the vdW forces between the crystal and the substrate \cite{huang2015reliable}. Next, the BSCCO flake is placed on a liquid nitrogen-cooled stage and cleaved into two pieces using a polydimethylsiloxane (PDMS) stamp at -$\SI{90}{\celsius}$. At this low temperature, the oxygen atoms in the BiO planes are frozen and no interfacial structural reconstruction occurs\,\cite{fratini2010scale}. The substrate is then quickly rotated to the desired angle and the flake on the PDMS is stacked on top of the other one. Less than 60 seconds elapse between the cleavage of the starting crystal and the assembly of the two parts. The stack is then warmed up to room temperature and released onto the substrate by gently lifting up the PDMS stamp. Immediately afterwards, we encapsulate the interface region with the hexagonal boron nitride thick layer via the PDMS transfer technique process in order to prevent the degradation at the junction resulting from chemical reactions with water or oxygen escape. Finally, electrical contacts are defined via stencil masks \cite{zhao2019sign} and deposited at -$\SI{30}{\celsius}$ in two steps in the evaporation chamber directly connected to the glovebox. Gold contacts are first established on the BSCCO flake, whereas the gold contacting pads are subsequently deposited with the chromium adhesion layer. To avoid local heating that causes loss of oxygen dopants the evaporation rate is set as low as fractions of $\SI{}{\angstrom\per\second}$. This implies that each deposition lasts around 3 hours, including the pumping and the venting of the chamber. The two steps evaporation aims at avoiding Cr directly onto the flakes since Cr oxidizes by ripping off oxygen molecules from the BSCCO. This undesired chemical reaction would lead to an insulating behavior of the region underlying the contact layer, strongly degrading the superconducting properties of the entire device\,\cite{ghosh2020demand}. Using this protocol, we obtain the high-quality electrical contact with the areal resistance smaller than $\SI{50}{\kilo\ohm\micro\meter^2}$, an order of magnitude lower than the contact resistance resulting from a single evaporation (Au/Cr) on BSCCO under our conditions. Figure\,\hyperref[fig:stacking]{1(b)} displays four representative twisted-JJs devices investigated in this study. All the flakes show a similar optical color contrast which corresponds to the thickness range of 30-$\SI{60}{\nano\meter}$. According to our experience and under our experimental conditions, below this thickness the JJ is of lower electronic quality or not entirely superconducting, possibly due to the lower robustness of each flake and due to the high spatial atomic modulation in thin crystals that affects the flatness of the interface.

To demonstrate the sharpness of our twisted interfaces at the atomic resolution, we perform cross-sectional high-annular dark-field scanning transmission electron microscopy in an extra heterostructure twisted at $\SI{45}{\degree}$. Bright spots in Fig.\,\ref{fig:TEM} correspond to electron scattering on atomic columns, the brightest of which are Bi that terminate each vdW layer and are the ones fully resolved. The crystalline order of each crystal is clearly preserved at the interface.

\begin{figure}[t!]
    \centering
    \includegraphics[width=.5\textwidth]{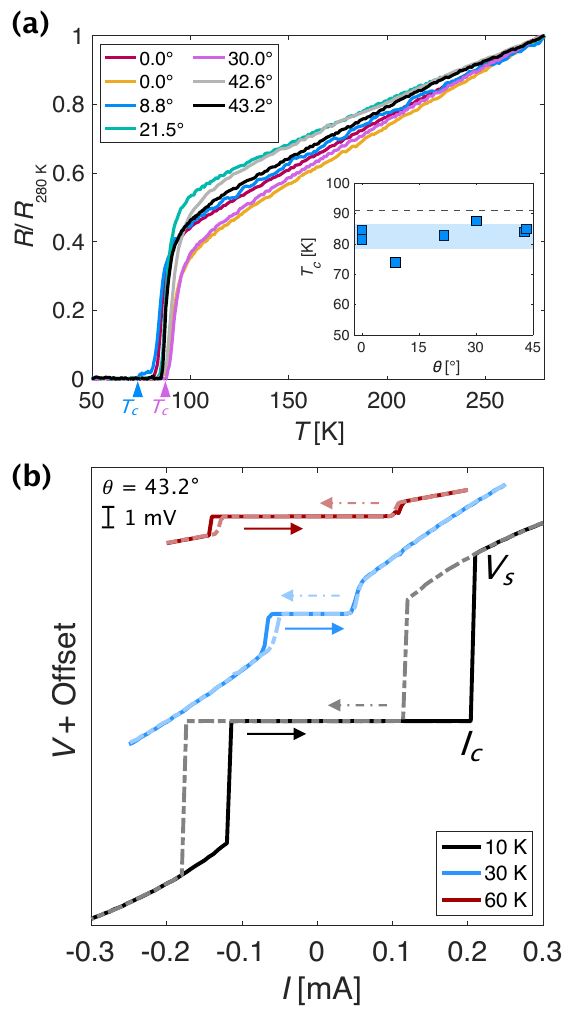}
    \caption{(a) Temperature-dependent electrical resistance normalized at $\SI{280}{\kelvin}$ obtained through the interface of seven twisted BSCCO junctions. Superconducting critical temperature $T_{c}$ are marked by a blue and a pink triangle for $\SI{8.8}{\degree}$ and $\SI{30}{\degree}$ twisted JJs, respectively.} Inset of (a): Angle dependence of $T_{c}$ of corresponding the corresponding JJs. The blue-shaded area is around the mean value of the critical temperatures and its width is two times the standard deviation. The black dashed line indicates the $T_{c}$ of an optimally doped bulk BSCCO crystal. (b) Current-voltage ($I$-$V$) characteristics of a $\SI{43.2}{\degree}$-twisted JJ measured by sweeping the current in both directions (arrows) at $\SI{10}{\kelvin}$, $\SI{30}{\kelvin}$ and $\SI{60}{\kelvin}$.
    \label{fig:characterization}
\end{figure}

Figure\,\hyperref[fig:characterization]{3(a)} shows the electrical resistance across each of the seven junctions, measured injecting $\SI{1}{\micro\ampere}$-current in the four-terminal geometry as a function of temperature $T$. The temperature dependence of each resistance exhibits a linear behavior in the normal state, consistent with the optimally doped\,\biscco\,\cite{liao2018superconductor}. The $R(T)$ of the JJs devices is followed by the clear single superconducting (SC) transition occurring at the temperature slightly lower than the bulk critical temperature of $\SI{91}{\kelvin}$\,\cite{presland1991general}, indicating optimal oxygen doping even at the junction interface. The only exception is given by the JJ with $\theta =  \SI{8.8}{\degree}$, whose electrical resistance displays a tiny hump between $\SI{74}{\kelvin}$ and $\SI{80}{\kelvin}$, suggesting a slightly underdoped interface. The devices investigated in this study have an onset superconducting critical temperature $T_c$ between $\SI{74}{\kelvin}$ and $\SI{88}{\kelvin}$ with the average value of $\bar{T_c} = \SI{83}{\kelvin}$  [inset of \hyperref[fig:characterization]{3(a)}]. Through this study, $T_{c}$ refers indeed to the highest temperature at which the electrical resistance measured across the junction remains zero (marked by a triangle in Fig.~\hyperref[fig:characterization]{3(a)} for two representative JJs). We observe no correlation between $T_{c}$ and the twist angle. We ascribe instead the variability of $T_{c}$ to a different, albeit reduced, oxygen loss at the cleaved surface during the exfoliations, as far as the flakes are thick enough to guarantee a flat interface that allows coherent tunneling of Cooper pairs through the junction.  

The current-voltage ($I$-$V$) characteristics of a representative JJ with $\theta = \SI{43.2}{\degree}$ measured at three different temperatures are displayed in [Fig.~\hyperref[fig:characterization]{3(b)}]. The sharp SC transition across the JJs, especially visible at $\SI{10}{\kelvin}$, confirms significant switching at the interface. 
%The sharp SC transition across the JJs is also visible \textcolor{purple}{at $\SI{10}{\kelvin}$} in the $I$-$V$ characteristics [Fig.~\hyperref[fig:characterization]{3(b)}] of a representative JJ with $\theta = \SI{43.2}{\degree}$, confirming significant switching at the interface. 
As we sweep the electrical current from a large negative value, the junction voltage $V$ first retraps to the SC state ($V = 0$) and then jumps to the resistive state at the critical current $I_{\mathrm c}$. The switching voltage at $\SI{10}{\kelvin}$ is $V_{\mathrm S} = \SI{8.5}{\milli\volt}$. At low temperatures, the $I$-$V$ curve of the JJ exhibits the large hysteresis in the bias current sweep, directly associable to the dielectric nature of the tunnel barrier with the high capacitance in the underdamped regime\,\cite{barone1982physics}. By reversing the current polarity, we obtain the $I$-$V$ characteristic mirrored along the $I=0$ axis. As shown in Fig.~\hyperref[fig:characterization]{3(b)}, the hysteresis becomes weaker as the temperature is increased, in agreement with previous studies on intrinsic JJs. %\cite{}} %{For better visualization, we calculate the numerical derivative of the resistance $dV/dI$ across the corresponding JJ, shown in the inset of Fig.\,\hyperref[fig:characterization]{3(b)}.}

\section{Results \& Discussion}

 \begin{figure*}[t!]
    \center
    \includegraphics[width=.9\textwidth]{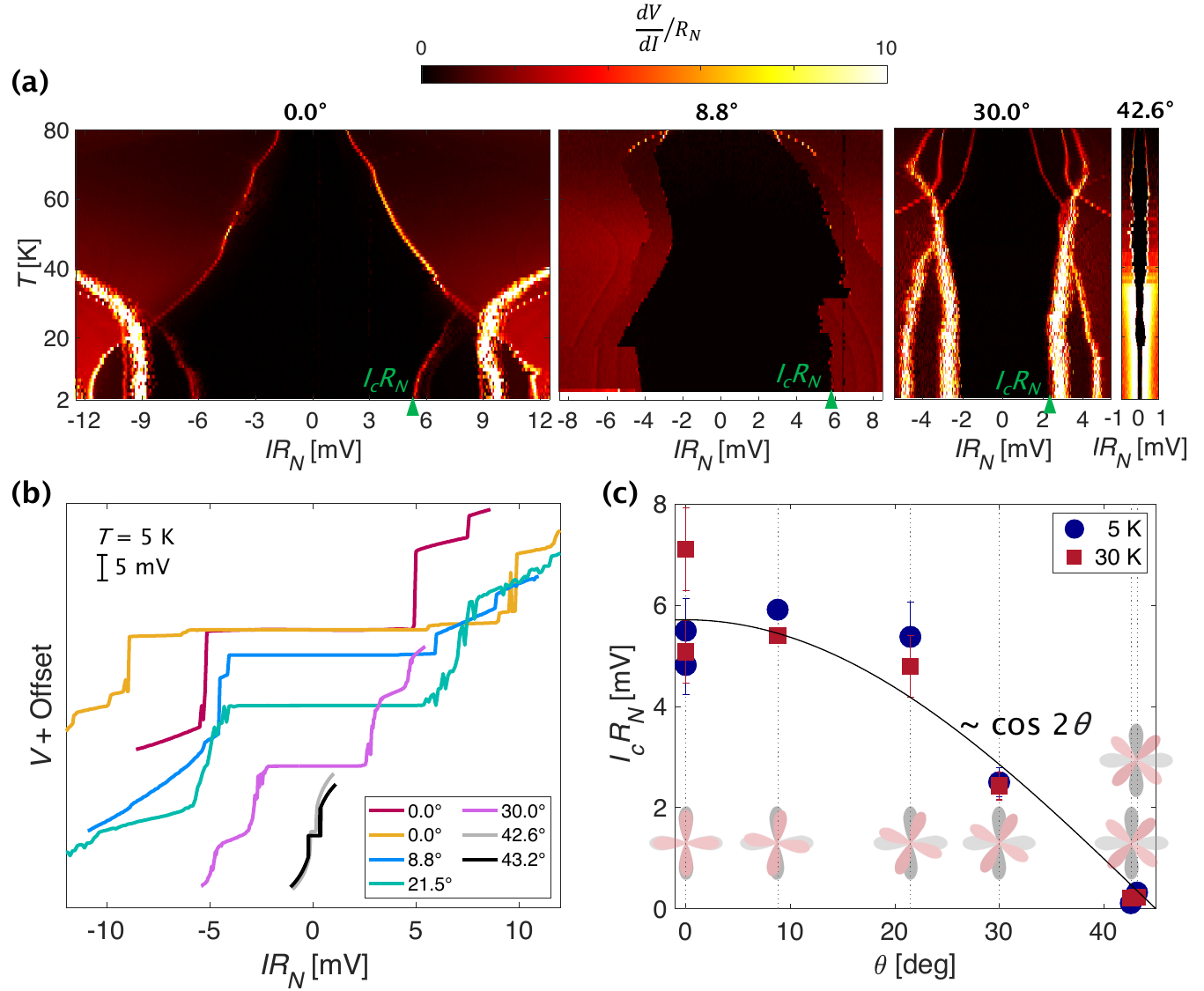}
    \caption{(a) Normalized differential resistance $[dV/dI]/R_{\text N}$ as a function of $IR_{\text N}$ and temperature $T$ from 2 to $\SI{80}{\kelvin}$ of four representative twisted JJs shown in Fig\,\hyperref[fig:stacking]{1(b)}. The twist angle is displayed on top of each color plot. The current is swept from large negative to positive bias. The green triangles highlight the position of $I_{\text c}R_{\text N}$ where the corresponding voltage jumps appear in the $IR_{\text N}$-$V$ curves. The JJ with $\theta = \SI{8.8}{\degree}$ shows a linear voltage with a small slope around zero bias above $\SI{74}{\kelvin}$. The corresponding $[dV/dI]/R_{\text N}$ is small, but finite, for low amplitudes of the injected current. (b) Normalized bias current-voltage ($IR_{\text N}$-$V$) characteristics for all seven JJs at \SI{5}{\kelvin}. Each curve is shifted along the y-axis for better visualization. (c) Angular dependence of $I_{\text c}R_{\text N}$ at $\SI{5}{\kelvin}$ and $\SI{30}{\kelvin}$. The solid line follows the $\cos{2\theta}$ curve, which is the angular dependence of the critical current in the first-  approximation for $d$-wave SOP. The error bars show the uncertainty on the value of $R_{\text N}$, extracted as the slope of $V(I)$ in the linear regime before switching the intrinsic junctions of the \biscco crystals (low bias). In correspondence with each marker, a schematic diagram of the $d$-wave wavefunctions of the crystals, twisted at the corresponding angle, is presented.
    \label{fig:Fig4}}
\end{figure*}

We investigate the pairing symmetry of the Cooper pairs in the twisted BSCCO JJs by measuring the $I$-$V$ characteristics and the dynamic resistance $dV/dI$ across the interface for several twisted JJs. We sweep a dc bias from a negative to positive value while superimposing
a tiny alternating current of $\SI{1}{\micro\ampere}$ amplitude with the frequency of $\SI{15}{\hertz}$, and measure simultaneously the dc and ac voltages across each junction. The details of the measurement are reported in the Supplementary Information. To compare the transport data of all the JJs, we normalize the bias current as $IR_{\text N}$, where $R_{\text N}$ is the junction resistance in the normal state, estimated as the slope of the $I$-$V$ curve in the linear region. 
Since both $I_{\text c}$ and $R_{\text N}^{-1}$ scale linearly with the junction area, their product is independent of it. Figure\,\hyperref[fig:Fig4]{4(a)} illustrates the normalized differential resistance $\frac{dV}{dI}/R_{\mathrm N}$ as a function of $IR_{\mathrm N}$ and temperature $T$ for four representative twisted JJs. Similar data for all seven JJs are included in Fig.\,S2 in the Supplementary Information (SI).  
The main feature that can be deduced from the color plots is that $I_{\text c}R_{\text N}$, at which the inner peaks of each $dV/dI$ appear, sensitively decreases at twist angles close to $\SI{45}{\degree}$. 

For better visualization, we show in Fig.\,\hyperref[fig:Fig4]{4(b)} the $I-V$ characteristics at $\SI{5}{\kelvin}$ in all the samples.
 For the two JJs with $\theta = \SI{0}{\degree}$, the critical currents normalized by the corresponding junction area are $j_{\mathrm c} = \SI{0.9}{\kilo\ampere\per\centi\meter^2}$ and $j_{\mathrm c} = \SI{1.0}{\kilo\ampere\per\centi\meter^2}$ at $\SI{10}{\kelvin}$ \cite{lee2022encapsulating}, comparable to $j_{\mathrm c}$ of intrinsic junction in BSCCO crystals, which ranges at that temperature from $\SI{0.17}{\kilo\ampere\per\centi\meter^2}$ to $\SI{1.7}{\kilo\ampere\per\centi\meter^2}$, depending on the number of junctions along the $c$-axis \cite{irie2000critical}. The realization of pristine interfaces allows us to rule out the effect of detrimental disorder at the junctions arising during the fabrication process, and to attribute any changes of $I_{\mathrm c}R_{\mathrm N}$ to the intrinsic effects of the BSCCO SOP symmetry. Differently from BSCCO JJs investigated in previous studies \cite{zhao2021emergent, lee2021twisted}, the $dV/dI$ at $\theta = \SI{0}{\degree}$ in Fig\,\hyperref[fig:Fig4]{4(a)} does not show hysteresis, suggesting a barrier with low capacitance. On the other hand, some of the JJs with finite twist angles show an asymmetric $dV/dI$ with respect to $IR_{\mathrm N} = 0$, displaying a difference in amplitude between the critical current and the retrapping current, appearing in the underdamped region, see Fig.\,S2 in the SI. The amount of damping is generally insensitive to the angle but it depends on the dielectric nature of the junction's capacitance. Nevertheless, we observe no correlations, for instance, between the tunneling property (damping regime) and the time between cleaving and stacking the flakes to create the interface, which is comparable for all devices and in all the cases below $\SI{60}{\second}$. One possible factor that determines the capacitance of the barrier is the purity level of argon in the glovebox which could change during the fabrication of the different interfaces. As we do not observe water condensation down to $-\SI{95}{\celsius}$, we estimate the partial pressure of water in the order of 10 ppb. This extremely low density of water molecules might slightly vary depending on the humidity condition of the laboratory due to finite exchange with the glovebox through the butyl gloves. Another factor that could give rise to a difference in the tunnel barrier's capacitance between the samples is the way the top flake stacks on top of the bottom one within the interface region. Two other important features are visible in the set of the $dV/dI$ color maps. In the twisted JJs with $\theta>\SI{20}{\degree}$, the $I_{\mathrm c}R_{\mathrm N}$ has a nonmonotonic temperature dependence: it increases above $\SI{40}{\kelvin}$, before dropping close to $T_{\mathrm c}$. This behavior can be related to the extrinsic mechanisms active at higher temperatures and/or to the intrinsic competition between the supercurrent contributions from nodal and anti-nodal regions of the Fermi surface driven by tunnel directionality\,\cite{gabovich2014anomalous}. 
 
 In the high-temperature regime, the tunneling becomes less coherent at the two contributions that have different magnitudes \cite{tummuru2022josephson}, resulting in a higher net $I_{\mathrm c}$. Secondly, additional peaks in all $dV/dI$ appear at bias exceeding the critical current, which can be associated either with the multiparticle tunneling\,\cite{schrieffer1963two}, such as phonon-assisted tunneling process\,\cite{schlenga1998tunneling}, and to intrinsic JJs intruding the current path between the voltage leads\,\cite{kim1999suppressed}. 
  
 Figure \hyperref[fig:Fig4]{4(c)} displays the angular dependence of $I_{\mathrm c}R_{\mathrm N}$ at two representative temperatures $\SI{5}{\kelvin}$ and $\SI{30}{\kelvin}$. We notice that $I_{\mathrm c}R_{\mathrm N}$ follows a $\cos{2\theta}$ dependence, which is the expected behavior of the critical current for tunneling between the $d$-wave superconductors in the first order approximation. The values of $I_{\mathrm c}R_{\mathrm N}$ in the two JJs with a tilt angle close to $\SI{45}{\degree}$ are about two orders of magnitude lower than that in the slightly twisted JJs. The critical current is significantly reduced in these two JJs due to the nearly maximum mismatch between the $d$-wave SOPs of the two crystals, but it still remains finite. This feature occurs, on one hand, since the tilt angle is not exactly $\SI{45}{\degree}$, but, on the other hand, even if it were, $I_{\mathrm c}R_{\mathrm N}$ should not completely vanish due to the nonzero contribution of the second harmonic component (at $\SI{45}{\degree}$) to the total critical current \cite{volkov2021josephson}.  
 
 Despite the different temperature profiles of $I_{\mathrm c}$ in the seven devices, the $\cos{2\theta}$ behavior of the angular dependence in $I_{\mathrm c}R_{\mathrm N}$ remains almost unchanged up to $\SI{80}{\kelvin}$. For $\theta = \SI{0}{\degree}$, $I_{\mathrm c}R_{\mathrm N}$ is around $\SI{6}{\milli\volt}$ at $\SI{10}{\kelvin}$, much smaller than the switching voltage amplitude of $V_{\mathrm S} = \SI{16}{\milli\volt}$ in the $I$-$V$ curve at the same temperature (not shown), which corresponds to the superconducting gap value in the ideal tunnel junctions\,\cite{barone1982physics}. This difference arises from the partial incoherent tunneling that does contribute to the conductivity ($\propto R_{\mathrm N}^{-1}$), but does not affect the $I_{\mathrm c}$ of a $d$-wave superconductor. As can be seen from the comparison between the switching voltage in Figs.\,\hyperref[fig:characterization]{3(b)} and the magnitude of $I_{\mathrm c}R_{\mathrm N}$ in Fig.\,\hyperref[fig:Fig4]{4(c)}, this discrepancy ($V_{\mathrm S} \neq I_{\mathrm c}R_{\mathrm N}$) becomes huge for $\theta = \SI{43.2}{\degree}$, since the $R_{\mathrm N}$ is not really dominated by the twist (especially the incoherent part), but $I_{\mathrm c}$ depends crucially on the symmetry of the Cooper pair wavefunction and direct Cooper-pair tunneling is expected to be strongly suppressed at $\SI{45}{\degree}$. \\

\section{Conclusions}

In summary, we adopt the cryogenic, solvent-free stacking technique to realize interfaces protected with an insulating hBN flake. We investigate the anisotropic SOP of BSCCO by studying the angular dependence of the critical current in the twisted JJs.  The superconducting transition of the twisted BSCCO junctions is shown to be very sharp as demonstrated by the temperature dependence of the resistance and the current-voltage characteristics, and occurs at the temperature comparable to the bulk value of $T_{\mathrm c}$. We find that the critical current density in the $\SI{0}{\degree}$-JJ is similar to that of the intrinsic JJs and drops by two orders of magnitude when the twist angle approaches $\SI{45}{\degree}$, demonstrating the $d$-wave superconducting order parameters. \\  

\noindent\textbf{Data availability}
The data that support the findings of this study are available from the corresponding author upon reasonable request.\\

\noindent\textbf{Acknowledgements.} The experiments were partially supported by the Deutsche Forschungsgemeinschaft (DFG 452128813, DFG 512734967, DFG 492704387, DFG 460444718). D.W. acknowledges funding from DFG SFB 1415, Project ID No.  41759051. The work of V.M.V. was supported by the Terra Quantum AG. K.W. and T.T. acknowledge support from the JSPS KAKENHI (Grant Numbers 19H05790, 20H00354 and 21H05233). The work at BNL was supported by the US Department of Energy, oﬃce of Basic Energy Sciences, contract no. DOE-sc0012704. The authors are grateful to Heiko Reith and Nicolas Perez for providing access to cleanroom and cryogenic facilities respectively. The authors thank Thomas Wiek and Dina Bieberstein for TEM sample preparation. The authors are also  grateful to S. Y. Frank  Zhao, Philip Kim, Uri Vool and Valentina Brosco for illuminating and fruitful discussions.\\

\noindent\textbf{Author contributions.} N.P. conceived and designed the experiment; M.M., Y.L., T.C. performed the experiments and analyzed the data. The cuprate crystals have been provided by G.G. The hexagonal boron nitride crystals have been provided by K.W. and T.T.  The fabrication procedure and the results have been discussed by N.P., M.M., Y.L. and  V.M.V. The manuscript has been written by N.P., M.M., Y.L., V.M.V., and K.N. All authors discussed the manuscript.\\

\noindent\textbf{Declaration of Interest} All authors declare no conflict of interest.

\bibliographystyle{}

%\FloatBarrier
%\bibliography{bibliography}
%\bibliographystyle{ieeetr}

\end{document}